# Measuring Tune, Chromaticity and Coupling

*Rhodri Jones*
CERN, Geneva, Switzerland

**Abstract**
This chapter takes a look at the ways tune, chromaticity and coupling can be measured in synchrotrons. After briefly introducing the importance of these parameters for machine operation, a broad overview of the various instrumentation and analysis techniques used in their determination will be given.

## 1 Introduction

The instrumentation used to observe transverse beam motion is very important for the efficient operation of any circular accelerator. There are three main parameters that can be determined using such diagnostics, namely the betatron tune, chromaticity and betatron coupling, all of which are discussed in detail in this chapter.

## 2 Betatron Tune

The betatron tune is a characteristic of the magnetic lattice and is defined, to first order, by the strength of quadrupole magnets. It can be thought of as the number of oscillations a particle which is not on the central orbit (defined by the centre of all quadrupoles) will undergo while completing a full revolution. The full betatron tune, Q, can be split into two components: the integer tune, defined as the number of complete oscillations the particle undergoes during one revolution, and the fractional tune, q, representing the fractional difference in the phase of the oscillation from one turn to the next. It is typically this fractional part of the tune that is measured to find the optimal working point for the accelerator.

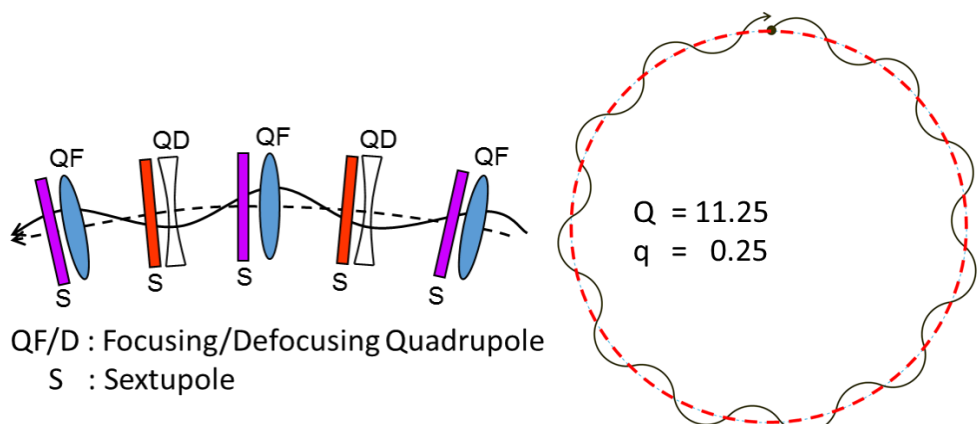

**Fig. 1:** The betatron tune is the characteristic frequency of the magnetic lattice and can be represented by an integer part and a fractional part

The integer tune is typically measured by looking at the residual betatron pattern imprinted on the global beam orbit using the accelerator's beam position measurement system, while the fractional part of the tune, which is typically the quantity of interest for beam optimisation, is usually determined using measurements from a specific monitor installed at one location in the machine.

With a single particle in the machine, a sensor at one specific location will see this particle passing at the revolution frequency, giving a series of output pulses (denoted here by the Dirac delta function δ) at intervals equal to the revolution period, $T_{rev}$. A Fourier transform of this signal, $\mathcal{F}[x(t)]$, gives the picture of this pattern in the frequency domain, which results in a series of lines at all multiples of the revolution frequency, $\omega_{rev} = 2\pi f_{rev}$.

$$x(t) = \sum_{n=-\infty}^{\infty} \delta(t - nT_{rev}) \qquad \mathcal{F}[x(t)] = \sum_{k=-\infty}^{\infty} \omega_{rev}\, \delta(\omega - k\omega_{rev}) \qquad (1)$$

If we replace the single particle with an ensemble of particles, $y(t)$, having a Gaussian distribution with standard deviation $\sigma$,

$$y(t) = \frac{1}{\sqrt{2\pi}\sigma} e^{-\frac{t^2}{2\sigma^2}} \qquad \mathcal{F}[y(t)] = e^{-\frac{\omega^2}{2(1/\sigma)^2}} \qquad (2)$$

then the time domain representation (Fig 2(a)) becomes a series of Gaussian pulses, which is simply the convolution of the two functions, $x(t) * y(t)$. In the frequency domain, by the convolution theorem,

$$\mathcal{F}[x(t) * y(t)] = \mathcal{F}[x(t)] \times \mathcal{F}[y(t)] = e^{-\frac{\omega^2}{2(1/\sigma)^2}} \sum_k \omega_{rev}\, \delta(\omega - k\omega_{rev}) \qquad (3)$$

The resulting picture in frequency domain is again a series of lines at all multiples of the revolution frequency, but now with a Gaussian amplitude profile that is a function of the bunch length (Fig. 2(b)).

When a single, Gaussian bunch performs betatron oscillations then the turn-by-turn intensity signal from a position sensitive sensor will fluctuate at a frequency $\omega_q = 2\pi q f_{rev}$ with an amplitude A, where A < 1 (Fig. 2(c)). This can be described by the function

$$z(t) = y(t) * \left[ (1 + A\cos(\omega_q t)) \sum_{n=-\infty}^{\infty} \delta(t - nT_{rev}) \right]$$

$$= y(t) * \left[ \sum_{n=-\infty}^{\infty} \delta(t - nT_{rev}) + \frac{1}{2}A(e^{i\omega_q t} + e^{-i\omega_q t}) \sum_{n=-\infty}^{\infty} \delta(t - nT_{rev}) \right]$$

$$= y(t) * \left[ x(t) + \frac{1}{2}A e^{i\omega_q t} x(t) + \frac{1}{2}A e^{-i\omega_q t} x(t) \right]$$

From the properties of Fourier transforms it follows that

$$\mathcal{F}[z(t)] = \mathcal{F}[y(t)] \times \mathcal{F}\left[ x(t) + \frac{1}{2}e^{i\omega_q t} x(t) + \frac{1}{2}e^{-i\omega_q t} x(t) \right]$$

$$= \mathcal{F}[y(t)] \times \left[ \mathcal{F}[x(t)] + \frac{A}{2}\mathcal{F}[e^{i\omega_q t} x(t)] + \frac{A}{2}\mathcal{F}[e^{-i\omega_q t} x(t)] \right]$$

$$= e^{-\frac{\omega^2}{2(1/\sigma)^2}} \left[ \sum_{k=-\infty}^{\infty} \omega_{rev}\, \delta(\omega - k\omega_{rev}) \right]$$

$$+ \frac{A}{2} e^{-\frac{\omega^2}{2(1/\sigma)^2}} \left[ \sum_{k=-\infty}^{\infty} \omega_{rev}\, \delta\big((\omega - \omega_q) - k\omega_{rev}\big) + \sum_{k=-\infty}^{\infty} \omega_{rev}\, \delta\big((\omega + \omega_q) - k\omega_{rev}\big) \right]$$

The result is a series of revolution harmonics with sidebands at $\omega \pm \omega_q$. The envelope function for the revolution harmonics is again a Gaussian function dependent on the bunch length, while the envelope of the sidebands is a Gaussian that depends on both bunch length and modulation depth (Fig. 2(d)).

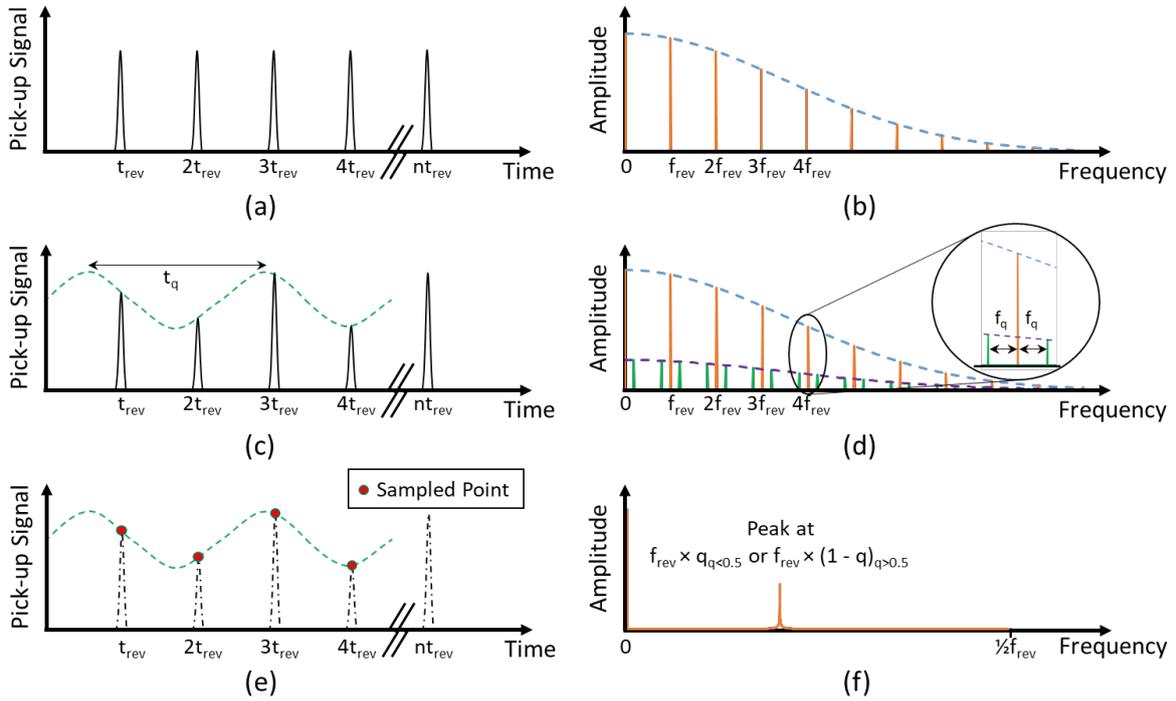

**Fig. 2:** Time and frequency domain representation for a bunch of particles observed at one single location on the circumference of the accelerator. (a & b) continuous measurement without betatron oscillation; (c & d) continuous measurement undergoing betatron oscillation (50% modulation); (e & f) sampled once per revolution.

As the original signal is sampled at the revolution frequency (one measurement per turn, see Fig. 2(e)), the Nyquist-Shannon theorem dictates that the maximum frequency for which the signal can be unambiguously reconstructed is $0.5 \times f_{rev}$. Beyond this frequency aliasing occurs, with these higher frequencies appearing as $(1 - \frac{f_{>0.5}}{f_{rev}}) \times f_{rev}$ (Fig. 2(f)). This means that from such a measurement it is not possible to determine whether the tune lies in the range $0 \rightarrow 0.5 \times f_{rev}$ or $0.5 \times f_{rev} \rightarrow f_{rev}$. The frequency spectrum for a tune of 0.2 will look identical to the frequency spectrum for a tune of 0.8. In a real machine the only way to verify in which range the tune lies is to vary the tune and look at how the spectrum changes. An increase in tune for a tune of 0.2 will lead to the sidebands moving further away from the revolution harmonics, while an increase for a tune of 0.8 will move the sidebands closer to the revolution harmonics.

## 2.1 Measuring Betatron Tune

Measuring and optimising the betatron tune through the whole operational cycle of a circular accelerator is one of the most important and basic control room tasks, and strongly influences the beam quality and its lifetime. The integer tune is typically measured by looking at the residual betatron pattern imprinted on the global beam orbit using the accelerator's beam position system, while the fractional part of the tune, which is typically the quantity of interest for beam optimisation, is usually determined using measurements from a specific monitor installed at one location in the machine. Separate measurements are performed in order to obtain the fractional tune in both the horizontal and vertical plane that, for beam stability reasons, are never the same. The quality of the fractional tune measurement is very important for deriving other crucial beam parameters, such as chromaticity and betatron coupling (see later in this chapter).

The individual particles in a beam constantly undergo betatron oscillations, the envelope of which define the beam size at any location (Fig. 3 (a)). As the phase of all particles is random, the motion is incoherent, with the beam centroid invariant. A beam measurement device that can only measure the centre of charge, such as a beam position monitor (BPM), will therefore observe a constant signal proportional to the total intensity and average position. In order to achieve coherent motion, external excitation is required to impose the same phase on all particles. If there are no energy loss mechanisms involved, which is true to first order for hadron beams, these particles will then continue to oscillate at this new amplitude. Any difference in their frequencies, however, will ultimately result in incoherent motion (Fig 3. (b)). The centre of charge of such an excited beam will therefore decohere with time, while the beam size increases to encompass the new oscillation amplitude. This sets the limitation for tune measurement techniques, which need to minimise the excitation required to observe a coherent signal in order to preserve the beam size.

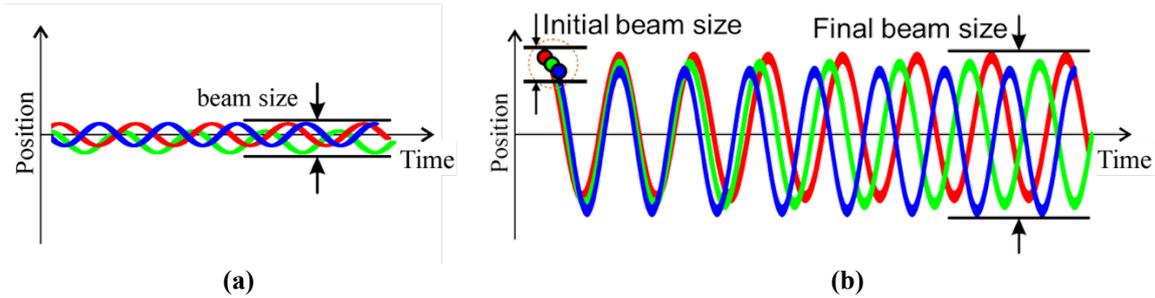

(a)                   (b)

**Fig. 3:** (a) The beam size is given by the envelope of the residual, incoherent betatron motion of all particles. (b) Applying external excitation imposes an initial common phase on all particles, which then oscillate coherently at the betatron tune, with the centre of charge decohering as a function of the difference in their frequencies (the tune spread).

### 2.1.1 *Tune measurement systems*

A tune measurement system can be based upon turn-by-turn readings from a regular beam position acquisition system (see Figs. 4, 5 6). Performing a Fast Fourier Transform (FFT) on this data will result in spectra showing the oscillation frequencies present in the transverse beam motion.

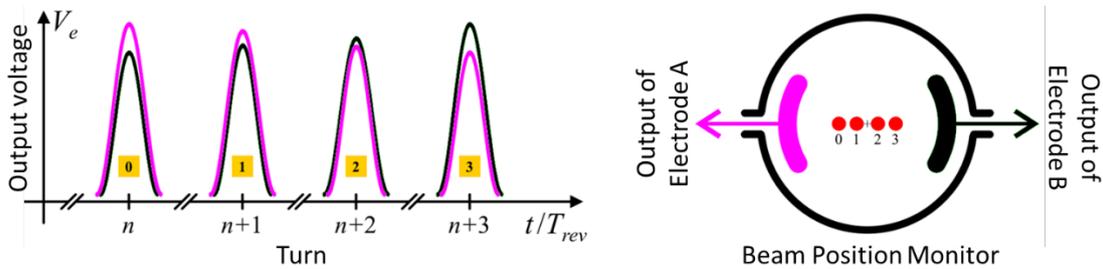

**Fig. 4:** Detecting oscillations using a beam position monitor. The oscillation information is superimposed as a small modulation on a large intensity signal.

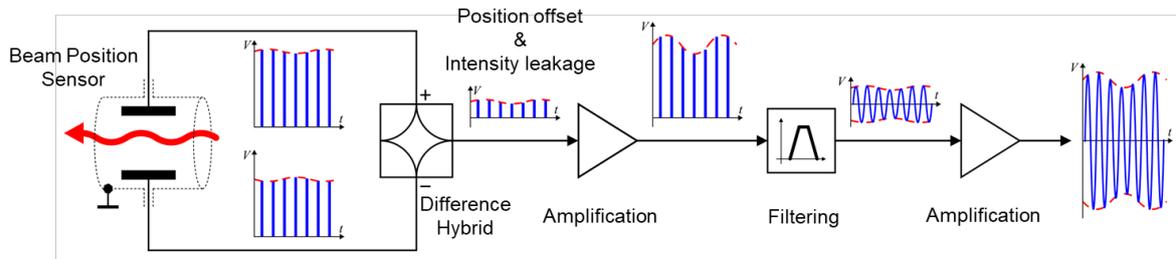

**Fig. 5:** Using a typical beam position measurement system for detecting oscillations. Most of the dynamic range of the system is used to determine the static position offset (the average amplitude of the resulting signal), with very little of the final signal actually corresponding to the oscillations of interest.

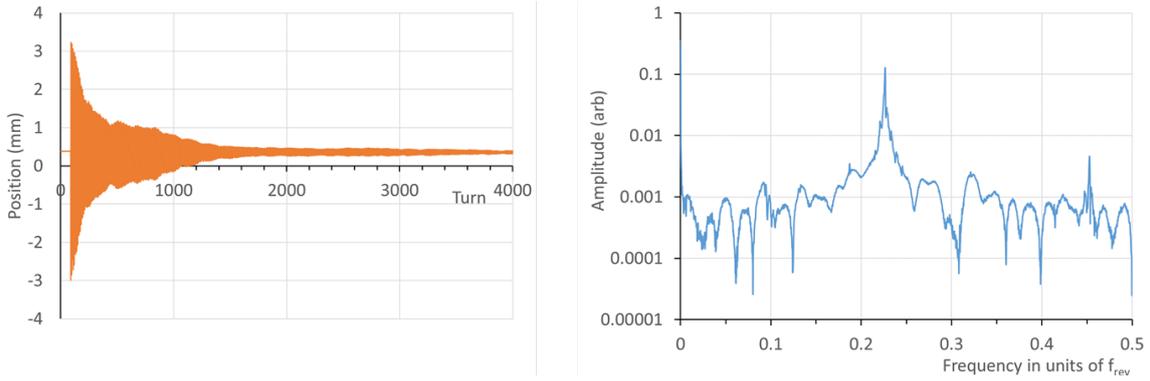

**Fig. 6:** Single kick tune measurement example from the CERN-SPS. Left – turn-by-turn position data from a beam position monitor. Right – the FFT of this data clearly showing the tune peak at 0.23.

Such a universal system is in many cases sufficient, especially for machines that can tolerate relatively large beam excitation (such as electron machines where radiation damping counters any emittance blow-up introduced by exciting the beam for a short period, or fast cycling machines where reduced beam quality is acceptable for a single cycle). Its main limitation is sensitivity, originating from the fact that the BPM system is optimised for accurate position measurements over a large dynamic range in both beam position and intensity. For example, if the system has to measure position over ±10 mm, then to detect oscillations with a 1 μm amplitude requires a dynamic range of 20000 on a turn-by-turn basis. Add to this an intensity range that can vary by an order of magnitude or more and the dynamic range required comes close to $10^6$. The intensity dependence is often taken care of by adapting the initial gain stage to the type of beam being measured, but this implies added complexity, with a prior knowledge of the beam characteristics required before launching a measurement. In the example of Fig. 6 the 3mm kick used results in a signal-to-noise ratio of ~100 at the tune peak in the FFT. This could easily be reduced by an order of magnitude while still achieving an acceptable signal to noise ratio at the tune peak, but going much beyond this, i.e. below ~100 μm, would start to affect the tune measurement resolution.

Tune measurements can be significantly improved by building dedicated systems optimised for beam oscillation detection rather than accurate beam position determination. In these systems the static beam position is rejected at a very early stage with only the oscillation signal retained for further processing. One such system is the so called Base-Band Tune (BBQ) system, based on a direct diode detection method initially designed for the LHC [1]. This allows a reliable tune measurement with micrometre or even sub-micrometre beam oscillations. In many machines such a level of coherent beam oscillation is always present, allowing measurement of the betatron tune without the need for additional excitation.

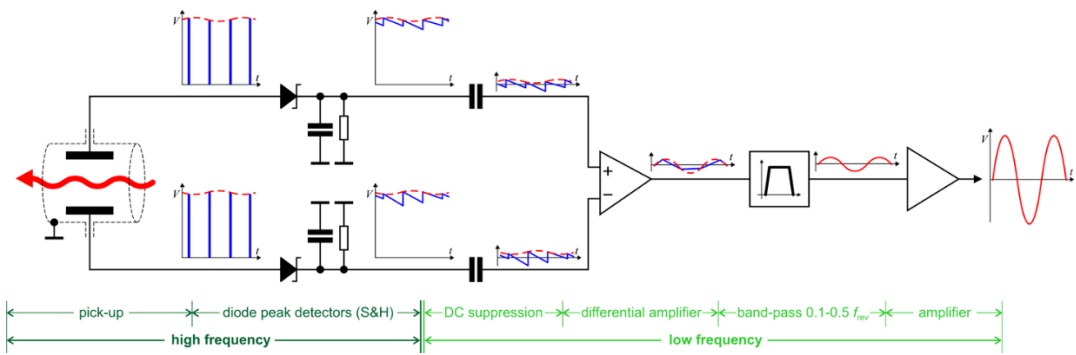

**Fig. 7:** Principle of operation of a base-band tune measurement system using direct diode detection.

In this technique the electrode signals from a beam position monitor undergo envelope demodulation with diode peak detectors, which can be considered as a simple, fast, sample-and-hold circuit (see Fig. 7). Most of the pick-up signal, related to the intensity and average beam position, does not change from one turn to another and gets converted to a DC level, while the small transverse beam oscillation is kept as a modulation on this DC offset. A series capacitor blocks the large DC component, while allowing the small modulation signal to pass. As the beam oscillation modulation from the opposing electrode is of opposite phase, enhancement of the signal is possible by subtraction using a differential amplifier. This has the added benefit of supressing common mode interference. The tune range of interest corresponds to half the machine revolution frequency, ranging from tens of kHz to a few MHz, depending on the size of the accelerator. The beam oscillations of importance after the diode peak detectors are therefore easily filtered, amplified and digitised with high resolution ADCs.

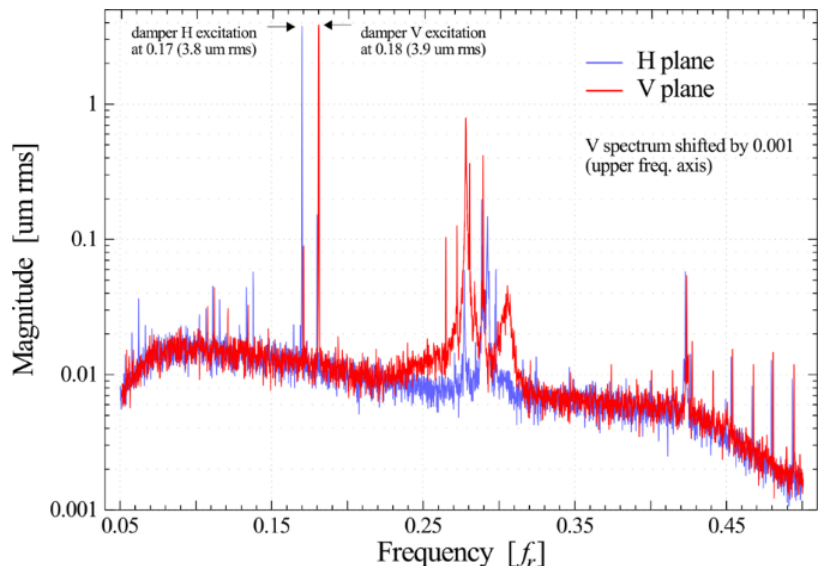

**Fig. 8:** Base band tune spectrum from the LHC. The residual horizontal and vertical tune oscillations at around 0.28 - 0.30 × $f_{rev}$ can be compared to forced excitation at known amplitudes at 0.17 × $f_{rev}$ and 0.18 × $f_{rev}$, showing a noise floor at the tens of nanometre level.

An example of a tune measurement acquired with such a BBQ device in the LHC is shown in Fig. 8. The FFT was acquired while intentionally exciting the beam in both the horizontal and vertical plane with a constant frequency that was visible as measurable oscillations by the standard beam position system. This allowed a calibration of the excitation amplitudes. The noise floor of the BBQ system can be seen to be at around the 10 nm level, while the residual beam oscillations at the tune frequencies are in the range of 50 nm to 800 nm. Coupling between the horizontal and vertical plane is also clearly visible.

The sensitivity of the BBQ system makes it ideal for continuous tune measurement in hadron machines as little or no additional excitation is required, so limiting emittance blow-up. Fig. 9(a) shows an example of a continuous horizontal tune measurement from the CERN SPS accelerator for a low intensity, single bunch. The horizontal tune is clearly visible at ~0.23 with a tune change applied later in the cycle easily tracked. Due to some coupling the vertical tune is also visible at ~0.16. Synchrotron sidebands are apparent at low frequency and around the horizontal tune with the decrease in the synchrotron frequency during the acceleration ramp observed as a reduction in their spacing. Fig. 9(b) shows a continuous measurement of the tune for the counter rotating LHC beams during an early commissioning ramp. At this time tune control was not optimised, with the tune crossing several resonance lines during acceleration. The loss in beam intensity correlated with these resonance crossings is clearly visible in the accompanying plot.

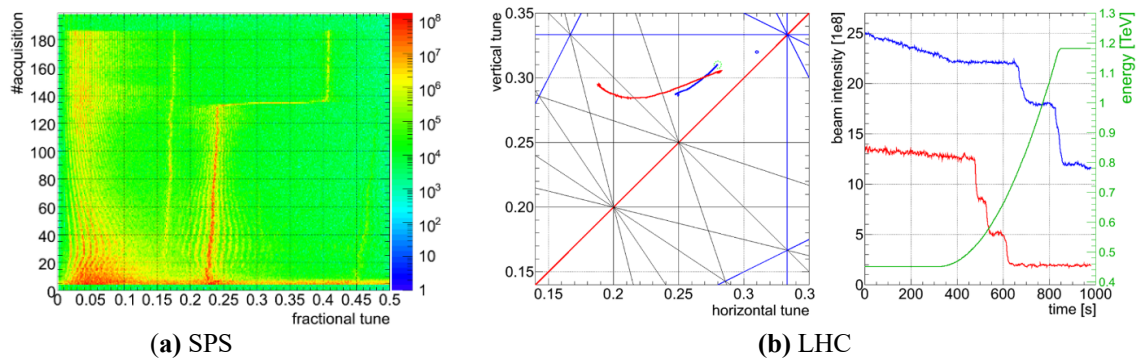

(a) SPS                               (b) LHC

**Fig. 9:** Examples of continuous tune measurement using the BBQ system. (a) Tune measurement with no additional external excitation for a low intensity single bunch in the CERN-SPS. (b) Horizontal and vertical tune for the two counter propagating beams measured during an early LHC ramp, along with the beam losses observed as the tunes cross resonance lines.

### 2.1.2 Tune measurement using a Phase Locked Loop

To make accurate tune measurements or fully characterise transverse dynamics, a Beam Transfer Function (BTF) measurement is often employed. This consists of exciting the beam with a steady sinusoidal signal and measuring the resulting beam motion at this specific frequency. The excitation frequency is then incremented by a small amount and a new measurement made. This process is repeated until the frequency range of interest is covered. With such a technique both the amplitude and relative phase of the beam oscillation can be precisely determined as the excitation frequency is known. A BTF measurement typically takes quite a long time, during which it must be assumed that all machine conditions remain constant. This is why the BTF method cannot be used for studying dynamic phenomena. An example of a beam transfer function is shown in Fig. 10. Notice how the phase jumps by 180° as the excitation is applied either side of the central betatron tune frequency. Such a response is typical for the forced excitation of any harmonic oscillator.

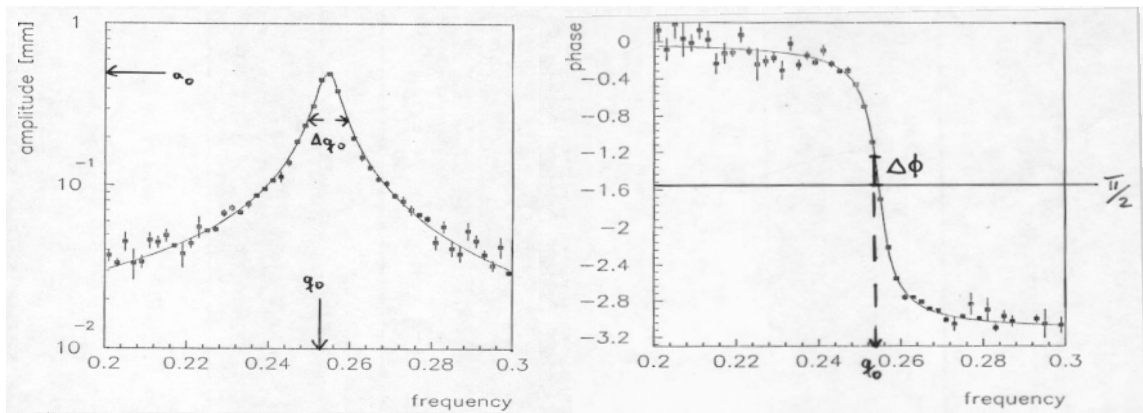

**Fig. 10:** Complete beam transfer function measured on the CERN-LEP collider.

This 180° phase jump is the basis for a Phase-Locked Loop (PLL) tune tracker. The principle of such a tracker is sketched in Fig. 11. A voltage or numerically controlled oscillator (VCO or NCO) is used to put a harmonic excitation $A \cdot \sin(\omega t)$ onto the beam. The beam response to this signal is then observed using a position pick-up. The response is of the form $B \cdot \sin(\omega t + \phi)$, where A and B are the amplitude of the excitation and oscillation respectively, $\omega$ is the angular excitation frequency and $\phi$ is the phase difference between the excitation and the observed beam response. In the phase detector both signals are multiplied, resulting in a signal of the form

$$\tfrac{1}{2} \cdot A \cdot B \cdot \cos(-\phi) - \tfrac{1}{2} \cdot A \cdot B \cdot \cos(2\omega t + \phi)$$

which has a DC component (first term in the equation) proportional to the cosine of the phase difference. This DC component is zero when the phase difference between excitation and observation frequencies is 90°, corresponding to the maximum amplitude response, i.e. at the tune frequency (see Fig. 10). The aim of the PLL is therefore to "lock-in" to this 90° phase difference between the excitation and observed signal by correcting the VCO frequency so that the DC component is always zero. The VCO frequency therefore tracks any tune changes, resulting in a continuous tune measurement.

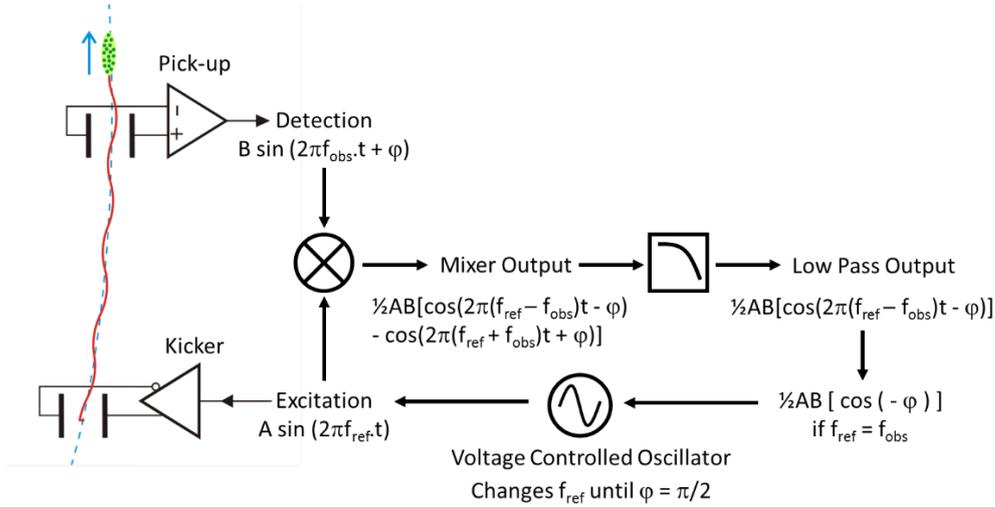

**Fig. 11:** Schematic layout of a phase locked loop tune tracker.

In practice the PLL tracker operation is influenced by many parameters, which have to be optimised so that the PLL finds, locks-in and subsequently tracks any tune changes. The beam spectra and dynamics have to be well understood if the PLL is not to lock or jump to a synchrotron sideband or an interference line for example. In addition, for hadron machines, the continuous excitation required may lead to emittance blow-up if sensitive acquisition electronics are not used.

## 3    Chromaticity

Chromaticity is defined as the difference in a particle's tune from the nominal tune due to a difference in its energy from the nominal energy. It is a function of the machine lattice and can be controlled by sextupole magnets. In a light-optics analogy, chromaticity can be thought of as the variation in focal length for different wavelengths when passing through a lens. In the beam case, the lens is a quadrupole magnet and the different wavelengths correspond to particles of different energy. For a particle with a momentum offset δp with respect to the nominal momentum p the difference in its tune, δQ with respect to its nominal tune Q is given by

$$\delta Q = Q' \frac{\delta p}{p}$$

where Q' is the chromaticity (also sometimes expressed as ξ = Q' / Q ). For a beam composed of many particles each with a different energy offset this leads to an energy spread, which in turn leads to a tune spread that is dependent on chromaticity.

For any high energy synchrotron, the control of chromaticity is very important. If the chromaticity is of the wrong sign (positive below the transition energy or negative above transition) then the beam quickly becomes unstable due to the head-tail instability. If the chromaticity is too large then the tune spread becomes large and some particles are inevitably lost as they hit resonance lines in tune space.

**3.1 Measuring Chromaticity**

There are many ways of measuring chromaticity, with the most common techniques summarised in Table 1. A few of these possible methods will be discussed in more detail in the following sections.

**Table 1:** Chromaticity measurement techniques

| Measurement Technique | Comments |
|---|---|
| Tune change for different beam momenta | Standard method used on most machines. Can be combined with tune tracking to give on-line measurements. |
| Width of tune peak or damping time | Model dependent and sensitive to non-linear effects. Not compatible with active transverse damping. |
| Amplitude ratio of synchrotron sidebands | Difficult to exploit in hadron machines with low synchrotron tune. Can be influenced by collective effects. |
| Width ratio of Schottky sidebands | Ideally suited to ion and un-bunched beams. The need for significant averaging makes this measurement very slow. |
| Bunch spectrum variations during betatron oscillations | Difficult to disentangle effects from other sources such as the bunch filling pattern, pick-up response and electronics noise. |
| Head-tail phase advance | Good results on several machines but requires kick stimulus which leads to emittance growth. |

*3.1.1 Chromaticity measurement by tune measurements at different beam momenta*

The standard technique of calculating chromaticity is to measure the betatron tune as a function of the beam energy and obtain the chromaticity from the resulting gradient. This is usually done by varying the RF frequency, keeping all magnetic fields static. The equations of interest are:

$$\Delta Q = (\xi Q)\frac{\Delta p}{p} = Q'\frac{\Delta p}{p} = Q'\gamma_t^2\frac{\Delta R}{R} = Q'\left(\frac{-\gamma_t^2\gamma^2}{\gamma^2 - \gamma_t^2}\right)\frac{\Delta f}{f}$$

where $\Delta Q$ is the change in tune, $\Delta p/p$ the momentum spread (or relative change in momentum), $\Delta R/R$ the relative change in radius, $\Delta f/f$ the relative change in RF frequency, $\gamma$ and $\gamma_t$ the relativistic $\gamma$ and $\gamma$ at transition respectively, and $\xi$ or $Q'$ (= $\xi Q$) the chromaticity.

In the CERN-SPS, for example, a chromaticity measurement consists of performing a tune measurement for three different RF frequency settings. Instead of noting the exact RF frequency and energy ($\propto \gamma$), what is actually measured is the change in closed orbit, from which the relative change in radius can be calculated. The three points are then plotted, with the gradient giving the chromaticity.

In order to obtain continuous chromaticity measurements this technique of RF modulation is combined with continuous tune measurement. As the RF modulation varies the momentum, the amplitude of the resulting tune modulation is directly proportional to the chromaticity. Examples from such chromaticity measurements at the LHC are shown in Fig. 12. The RF is varied by $< 10^{-3}$ at a frequency of $< 1$ Hz, to give orbit changes of 0.1 to 1 mm for the typical 1 m dispersion in the LHC arcs. A chromaticity resolution of 1 unit therefore requires a tune resolution of $< 10^{-4}$, something that can be achieved by demodulating at the correct frequency, averaging the tune measurements over many RF modulation periods, or adapting the PLL bandwidth. The final resolution obtained will be a trade-off between the RF modulation amplitude applied and the frequency or duration of the modulation.

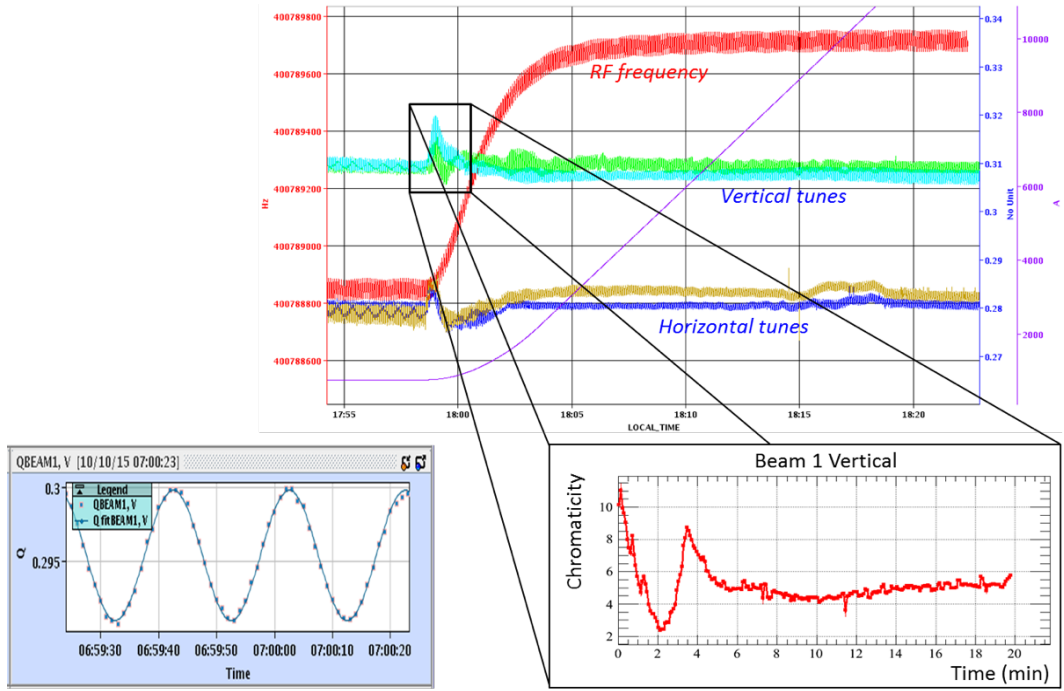

**Fig. 12:** Continuous chromaticity measurement in the LHC. The top plot shows the RF modulation (red) and the resulting tune modulation during the energy ramp. The chromaticity calculated from the tune modulation at the start of acceleration is shown at the bottom right. The plot on the bottom left shows the tune being tracked while the RF is modulated.

Instead of varying the frequency it is also possible to vary the RF phase. This typically allows for faster modulation, but may be limited by the RF power required. In this case the tune tracker does not need to follow the modulation but provides the carrier frequency for demodulation. The amplitude of this demodulated line is then a direct measure of the chromaticity [2, 3].

### 3.1.2 *Chromaticity measurement from synchrotron sidebands*

Due to the fact that the energy of a particle is modulated by its synchrotron motion, chromaticity will lead to tune changes that are modulated at the synchrotron frequency. This can be observed in the frequency domain as synchrotron sidebands appearing around the betatron tune peak. In the absence of space charge and collective effects, the ratio of the amplitude of these sidebands to the main tune peak is a direct measure of the chromaticity [4]. Such a technique has been used at the Diamond light source in the UK to measure chromaticity [5], where the RF modulation technique was not compatible with user operation (Fig. 13). A beam transfer function measurement (frequency sweep across the tune) was performed on a single bunch using the transverse bunch-by-bunch feedback system. The resulting tune spectrum, with its associated synchrotron sidebands, is then used to calculate chromaticity. The relationship between the sideband amplitude and chromaticity can be shown to have a Bessel function dependence.

Care must be taken when using such a technique with high intensity beams however, as the spectra can be severely modified by space charge and collective effects. Fits from analytical models are then required to extract both the chromaticity and space charge parameters [6].

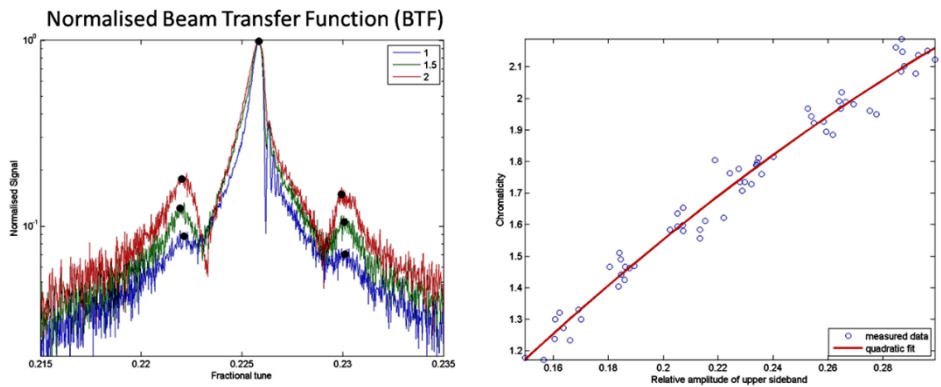

**Fig. 13:** Chromaticity measurement using the amplitude of the synchrotron sidebands at the Diamond light source, UK. The left plot shows the beam transfer function for 3 different chromaticities, while the right plot shows how the chromaticity is related to the sideband amplitude.

### *3.1.3    Chromaticity measurement from Schottky sidebands*

A Schottky diagnostic, looking at the incoherent motion of an ensemble of particles [7] is a powerful tool that, in principle, allows the non-invasive measurement of many accelerator parameters such as intensity, tune, chromaticity, momentum spread, synchrotron frequency and emittance. It is particularly effective for unbunched beams, where typical AC-coupled electromagnetic monitors cannot be used, and for ion beams where the signal scales with $Z^2$. Its success has been limited with bunched proton beams, where it is difficult to suppress the revolution harmonics to a high enough degree so as not to compromise the dynamic range of the acquisition system. Nevertheless several accelerators have managed to successfully use Schottky diagnostics for bunched proton beams, in particular to measure tune and chromaticity [8, 9, 10].

It can be shown [7] that the widths of the transverse Schottky sidebands around the $n^{th}$ revolution harmonic is given by

$$\Delta f_\pm = f_0 \frac{\Delta p}{p}[(n \pm q)\eta \pm Q\xi] \approx f_0 \frac{\Delta p}{p}[n \times \eta \pm Q'] \text{ for large } n$$

where $f_0$ is the revolution frequency, $\Delta f_\pm$ the upper (+) or lower (-) Schottky sideband, $\frac{\Delta p}{p}$ the momentum spread, $Q$ the full betatron tune, $q$ the fractional tune, $Q' = Q\xi$ the chromaticity and $\eta$ the machine parameter $1/\gamma_t^2 - 1/\gamma^2$ where $\gamma$ is the relativistic $\gamma$ and $\gamma_t$ that at transition energy.

From this equation it can be seen that the width of the upper and lower Schottky sideband around a given harmonic are not the same, and depend on chromaticity. In selecting the harmonic, *n*, to be used for a chromaticity measurement several constraints need to be taken into account:

1. For the best sensitivity to chromatic changes $n \times \eta$ should be of the same order as *Q'*, as this leads to the largest difference in width with chromaticity.

2. To avoid the Schottky sidebands from overlapping with the sidebands from neighbouring harmonics, $n < \left(4\eta \frac{\Delta p}{p}\right)^{-1}$

3. For bunched beams, *n* needs to be as high as possible so that the harmonic of choice is outside the single bunch spectrum such that the acquisition electronics is not saturated by revolution harmonics. For beams with a Gaussian longitudinal profile of width $\sigma_b$, one should aim for $n > \frac{6}{f_0}\left(\frac{1}{2\pi\sigma_b}\right)$ which should result in a factor $\sim 10^6$ reduction in the revolution line compared to that at low frequency. It should be noted that any high frequency components present in the longitudinal profile, resulting from a non-Gaussian profile, may imply significant power in the single bunch spectrum well beyond this harmonic.

The choice of harmonic is therefore a trade-off between all three of these constraints, aiming for the highest possible harmonic that avoids overlap, still provides reasonable sensitivity for chromaticity but is well above the longitudinal single bunch spectrum.

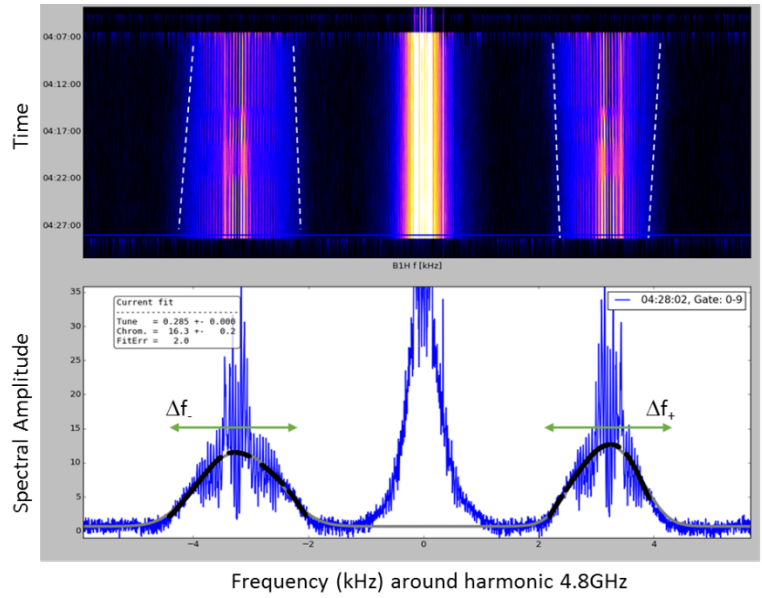

**Fig. 14:** Schottky chromaticity measurement from the LHC. The bottom plot shows a single Schottky spectrum centred on harmonic 436000 (4.8GHz). The top plot is a time evolution plot showing how this spectrum evolves as chromaticity changes.

An example of a bunched beam Schottky measurement in the LHC is shown in Fig. 14 for harmonic $n \sim 436000$, corresponding to 4.8 GHz ($f_0 = 11$ kHz). As can be seen, despite the harmonic being well above the presumed cut-off for a 0.3 ns sigma Gaussian longitudinal bunch profile (n > $\frac{6}{f_0}\left(\frac{1}{2\pi\sigma_b}\right)$ = 290000 ), there is still significant content in the revolution harmonic (central line in Fig. 14). The difference in the width of the two sidebands due to chromaticity is also clearly visible, as is the variation with time of the relative sideband widths as the chromaticity changes. The choice of harmonic is not optimal for chromaticity measurement, with $n \times \eta \sim 140$ implying that a one unit change in chromaticity represents only a 1.5% change in width. However, lowering the observation harmonic would have resulted in a much higher revolution line content, making it impossible to amplify the very low Schottky signals without saturating the electronics.

### 3.1.4   *Chromaticity measurement from head-tail motion*

Assuming longitudinal stability, a single particle will rotate in longitudinal phase-space at a frequency equal to the synchrotron frequency. During this longitudinal motion the particle also undergoes transverse motion. As the longitudinal motion (change in Δp/p) is coupled via chromaticity to the transverse motion (change in tune), this has the effect of phase modulating the betatron motion of a single particle as it performs its synchrotron oscillations. This is shown schematically in Fig. 15(a) where the phase modulation leads to an effective frequency modulation, the sign of which depends on whether the particle has more or less momentum than the synchronous particle.

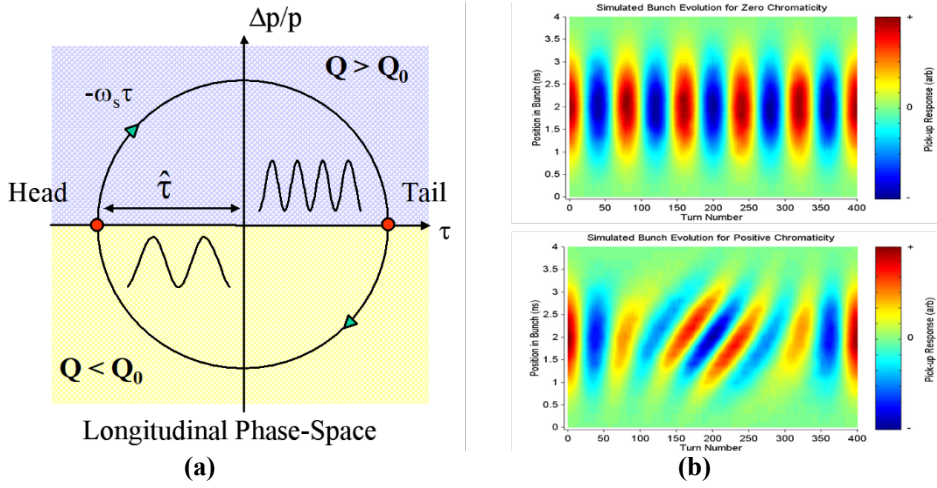

**Fig. 15:** (a) Schematic representation of the tune (betatron oscillations) of a single particle as it performs synchrotron oscillations. (b) Simulated response of an ensemble of particles in a single RF bucket to a transverse kick for zero chromaticity (top) and non-zero chromaticity (bottom) over a complete synchrotron period.

If the whole bunch is kicked transversely, then the resulting transverse oscillations for a given longitudinal position within the bunch can be shown [11] to be given by

$$y(n) = A\cos[2\pi n Q_0 + \omega_\xi \hat{\tau}(\cos(2\pi n Q_s) - 1)] \quad (1)$$

where $n$ is the number of turns since the kick, $Q_0$ is the betatron tune, $Q_S$ is the synchrotron tune, $\hat{\tau}$ is the longitudinal position with respect to the centre of the bunch, and $\omega_\xi$ is the chromatic frequency given by

$$\omega_\xi = Q'\omega_0 \frac{1}{\eta} \quad (2)$$

Here $Q'$ is the chromaticity, $\omega_0$ is the revolution frequency and η the machine parameter $1/\gamma_t^2 - 1/\gamma^2$ where γ is the relativistic γ, and $\gamma_t$ that at transition energy.

If we now consider the evolution of two longitudinal positions within a single bunch separated in time by Δτ, then from Eq. 1 it follows that the phase difference in the transverse oscillation of these two positions, ΔΨ, is given by

$$\Delta\Psi(n) = -\omega_\xi \Delta\tau(\cos(2\pi n Q_s) - 1) \quad (3)$$

This phase difference is a maximum when $nQ_S = ½$, i.e. after half a synchrotron period, giving

$$\Delta\Psi_{max} = -2\omega_\xi \Delta\tau \quad (4)$$

Such that the chromaticity can be written as

$$Q' = \frac{\eta \Delta\Psi_{max}}{2\omega_0 \Delta\tau} \quad (5)$$

By measuring the phase difference that develops on a turn by turn basis between the betatron oscillations of two longitudinal positions within a single bunch it is therefore possible to calculate the chromaticity. The effect of this variation in phase across the bunch is shown in Fig. 15(b), where the colour code represent the amplitude of oscillation (blue negative, green zero and red positive). For zero chromaticity the frequency of the betatron tune oscillation at any given location within the bunch is the same. However, with chromaticity the frequency for a given position within the bunch is modulated, leading to a phase difference that has a maximum at half the synchrotron period and whose value depends on chromaticity.

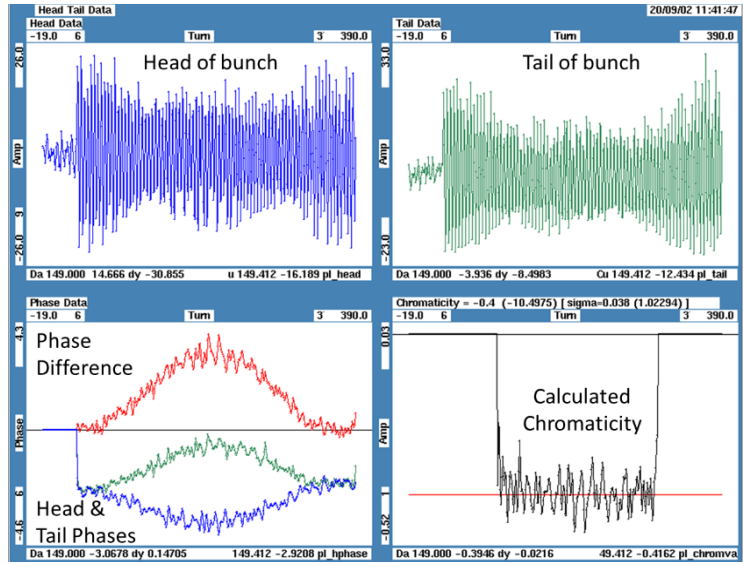

**Fig. 16:** Head-tail chromaticity measurement in the CERN-SPS. The top plots show the oscillation amplitude of a position at the head and tail of a single bunch after excitation using a single kick. The bottom left plot shows the phase modulation of the head and tail oscillations and the phase difference. The bottom right plot shows the chromaticity as calculated from the measured phase difference.

This so called "head-tail" chromaticity technique was developed as an alternative to RF modulation for fast chromaticity measurement (Fig. 16). However, despite being demonstrated at both the CERN-SPS [12] and HERA-p [13] it has up to now never been routinely used for operation due to several limitations. Firstly, the technique needs a strong kick to overcome limitations in the resolution of the electronic acquisition system, which can lead to emittance blow-up. It is also affected by space charge at low energy, which adds its own modulation to this motion. Lastly, it relies on decoherence times larger than the synchrotron period, which is not always the case. The technique should, in principle, also work with continuous excitation [14, 15], but as for the single kick case, for standard operation this requires sensitive electronics gated on the head and tail of the bunch, a development that has so far eluded beam instrumentalists.

## 4 Coupling

Coupling between the horizontal and vertical betatron motion in an accelerator is a result of skew magnetic components that have the effect of rotating the planes of oscillation. Horizontal motion is therefore visible in the vertical plane and vice-versa. As will be demonstrated in this section, the frequency of the observed oscillations in both planes will no longer be at the original horizontal and vertical betatron tune, but at frequencies modified by the extent of the coupling and the difference between the original horizontal and vertical tune.

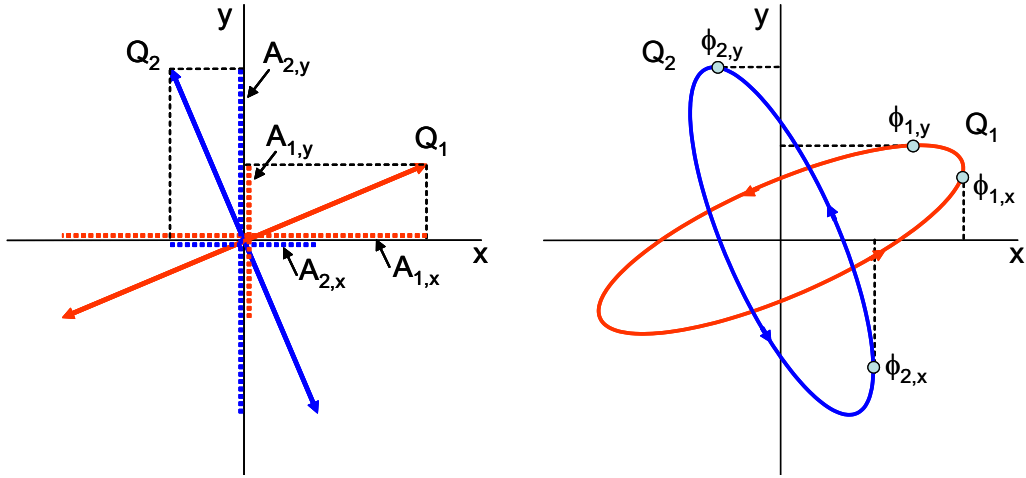

**Fig. 17:** Schematics showing the two eigenmodes rotated with respect to the horizontal and vertical planes due to coupling. The left hand figure shows the special case where the projections of each mode in each plane are in phase. The right hand side shows the more general case where coupling introduces a phase shift into the eigenmode projections. The measured tune corresponds to the frequency of the projected oscillation.

For a linearly coupled circular accelerator the observed displacement on turn n in the horizontal (x) and vertical (y) planes is a combination of the projections of two eigenmodes (see [16] and references therein). This is illustrated in Fig. 17 and can be expressed as

$$\begin{cases} x(n) = A_{1,x} \cos(2\pi Q_1 n + \phi_{1,x}) + A_{2,x} \cos(2\pi Q_2 n + \phi_{2,x}) \\ y(n) = A_{1,y} \cos(2\pi Q_1 n + \phi_{1,y}) + A_{2,y} \cos(2\pi Q_2 n + \phi_{2,y}) \end{cases} \quad (1)$$

Here it is assumed that Mode 1 is more linked to the horizontal plane, while Mode 2 is more linked to the vertical. The eigenmode frequency of Mode 1 is denoted by $Q_1$, while $A_{1,x}$ and $A_{1,y}$ represent the amplitudes of this mode in the horizontal and vertical plane respectively. Similarly $\phi_{1,x}$ and $\phi_{1,y}$ represent the corresponding phases of these modes. The same notation applies for the frequency, amplitudes and phases of Mode 2.

Using Hamiltonian perturbation theory in the absence of intentionally strong local couplers, it can be shown (see [16] and references therein) that the general expression for coupled betatron oscillations in the x and y plane along the reference trajectory can be written as

$$\begin{cases} x(s) = \sqrt{2\beta_x}\{a\cos[\Psi_x + (v-\Delta/2)\varphi - \chi/2] + b\cos[\Psi_x - (v+\Delta/2)\varphi - \chi/2]\} \\ y(s) = \sqrt{2\beta_y}\{c\cos[\Psi_y + (v+\Delta/2)\varphi + \chi/2] + d\cos[\Psi_y - (v-\Delta/2)\varphi + \chi/2]\} \end{cases} \quad (2)$$

where

$$\frac{c}{a} = -\frac{b}{d} = \frac{|C^-|}{2v+\Delta} \quad ; \quad v = \frac{1}{2}\sqrt{\Delta^2 + |C^-|^2} \quad ; \quad \varphi = \frac{2\pi s}{L}$$

Here s is the distance along the reference trajectory and L is the circumference of the accelerator. $\beta_x$ and $\beta_y$ are the optical betatron functions at s, $\Psi_x$ and $\Psi_y$ are the unperturbed horizontal and vertical angular frequencies, $\Delta$ is the difference between the fractional part of the unperturbed tunes ($\Delta = Q_{x,0} - Q_{y,0} - p$ with p an integer), and the complex coupling coefficient $C^-$ is defined as

$$C^- = |C^-|e^{i\chi}$$

When considering measurements taken at a single location on a turn-by-turn basis, Eq. (2) can be rewritten as

$$\begin{cases} x(n) = \sqrt{2\beta_x}\left\{ a\cos\left[2\pi(Q_{x,0} - \tfrac{1}{2}\Delta + \nu)n - \chi/2\right] + b\cos\left[2\pi(Q_{y,0} + \tfrac{1}{2}\Delta - \nu)n - (\pm\pi/2 + \chi/2)\right]\right\} \\ y(n) = \sqrt{2\beta_y}\left\{ c\cos\left[2\pi(Q_{x,0} - \tfrac{1}{2}\Delta + \nu)n + \chi/2\right] + d\cos\left[2\pi(Q_{y,0} + \tfrac{1}{2}\Delta - \nu)n + (\pm\pi/2 + \chi/2)\right]\right\} \end{cases} \quad (3)$$

Comparing Eq. (3) with Eq. (1), the following variables can be defined:

$$\begin{cases} r_1 = \dfrac{A_{1,y}}{A_{1,x}} = \sqrt{\dfrac{\beta_y}{\beta_x}} \cdot \dfrac{|C^-|}{2\nu + \Delta} \\[6pt] r_2 = \dfrac{A_{2,x}}{A_{2,y}} = \sqrt{\dfrac{\beta_x}{\beta_y}} \cdot \dfrac{|C^-|}{2\nu + \Delta} \end{cases} \quad (4)$$

$$\begin{cases} \Delta\phi_1 = \phi_{1,y} - \phi_{1,x} = \chi \\ \Delta\phi_2 = \phi_{2,x} - \phi_{2,y} = \pm\pi - \chi \end{cases} \quad (5)$$

It is also possible to write the following relations for the eigenmode frequencies, $Q_1$ and $Q_2$

$$\begin{cases} Q_1 = Q_{x,0} - \tfrac{1}{2}\Delta + \tfrac{1}{2}\sqrt{\Delta^2 + |C^-|^2} \\ Q_2 = Q_{y,0} + \tfrac{1}{2}\Delta - \tfrac{1}{2}\sqrt{\Delta^2 + |C^-|^2} \end{cases} \quad (6)$$

$$|Q_1 - Q_2| = \sqrt{\Delta^2 + |C^-|^2} \quad (7)$$

Solving for $\Delta$ and $|C^-|$ using Eq. (4) and (7) one obtains

$$|C^-| = \frac{2r_1\sqrt{\dfrac{\beta_y}{\beta_x}}|Q_1 - Q_2|}{\left(\dfrac{\beta_y}{\beta_x} + r_1^2\right)} \quad , \quad \Delta = \frac{|Q_1 - Q_2|\left(\dfrac{\beta_y}{\beta_x} - r_1^2\right)}{\left(\dfrac{\beta_y}{\beta_x} + r_1^2\right)} \quad (8)$$

From Eq. (4) it can be seen that $r_1 = \beta_y/\beta_x\, r_2$. Substituting into Eq. (8) one obtains the expressions for the coupling amplitude and the separation between the fractional part of the unperturbed tunes:

$$|C^-| = \frac{2\sqrt{r_1 r_2}\,|Q_1 - Q_2|}{(1 + r_1 r_2)} \quad , \quad \Delta = \frac{|Q_1 - Q_2|(1 - r_1 r_2)}{(1 + r_1 r_2)} \quad (9)$$

which are independent of the beta functions at the observation location.

### 4.1 Measuring Coupling

There are three main methods to measure the coupling in an accelerator: using orbit changes, looking at the closest tune approach or performing kick response measurements.

*4.1.1    Measuring coupling using orbit measurements*

To find coupling using orbit measurements, the orbit is changed in one plane by exciting steering correctors or by varying injection conditions, while the effect is measured in the other plane. Coupling sources can immediately be identified as locations where a horizontal orbit change generates a vertical kick and vice versa. By acquiring a large numbers of orbits while exciting different orbit correctors the skew quadrupole component of each magnet can be determined.

*4.1.2    Measuring coupling using the closest tune approach*

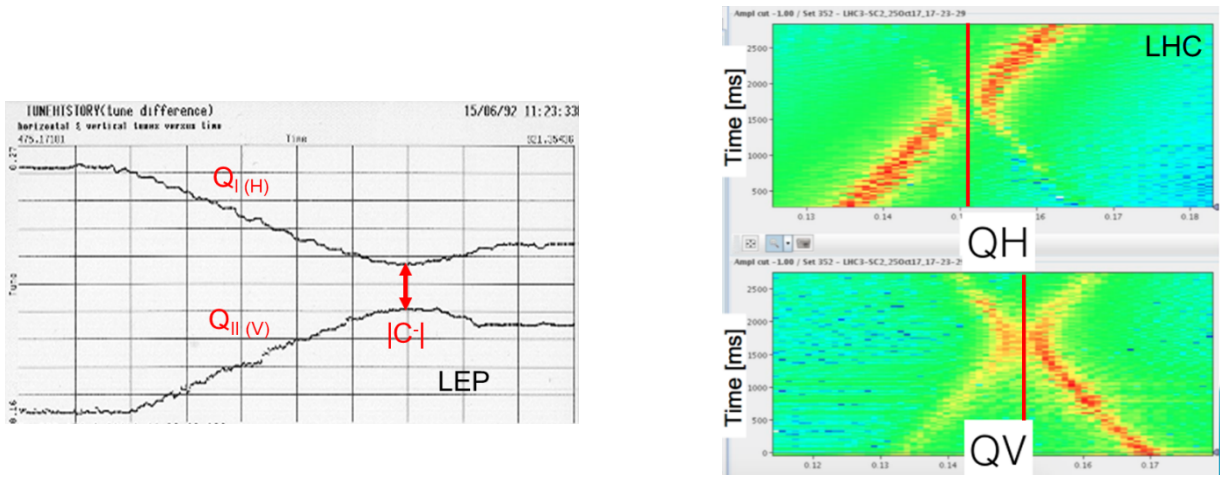

**Fig. 18:** Coupling measurements in LEP (left) and LHC (right) using the closest tune approach.

The closest tune approach relies on the fact that the frequency of coupled oscillators are modified depending on the coupling strength and their frequency difference. This is visible in Eq. (6) with the measured frequencies ($Q_1$ and $Q_2$) equal to the unperturbed frequencies ($Q_{x,0}$ and $Q_{y,0}$) plus a term that depends on the coupling and the difference between the unperturbed frequencies. When the two unperturbed frequencies are equal ($\Delta = 0$) then Eq. (7) states that the measured difference in the horizontal and vertical oscillations is equal to the magnitude of the coupling. By scanning one or both tunes across each other it is therefore relatively straightforward to measure the coupling (see Fig. 18).

*4.1.3    Measuring coupling using kick response*

There are several ways to use a kick response method to measure coupling. The simplest is to excite the beam with a single kick and measure the resulting oscillations using a tune measurement system at a given beam position monitor location. From Eq. (9) it can be seen that the coupling magnitude depends on the amplitude ratio of the horizontal and vertical oscillation frequencies and their separation, which is easily measured in the FFT of the oscillation data (Fig. 19).

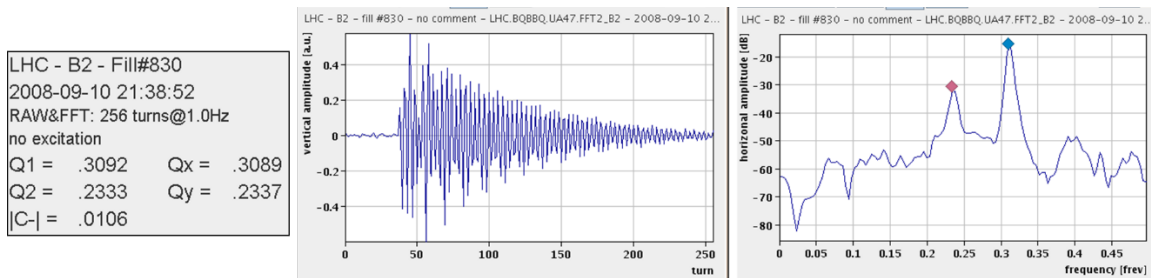

**Fig. 19:** Coupling calculated in the LHC using the response to single kick excitation.

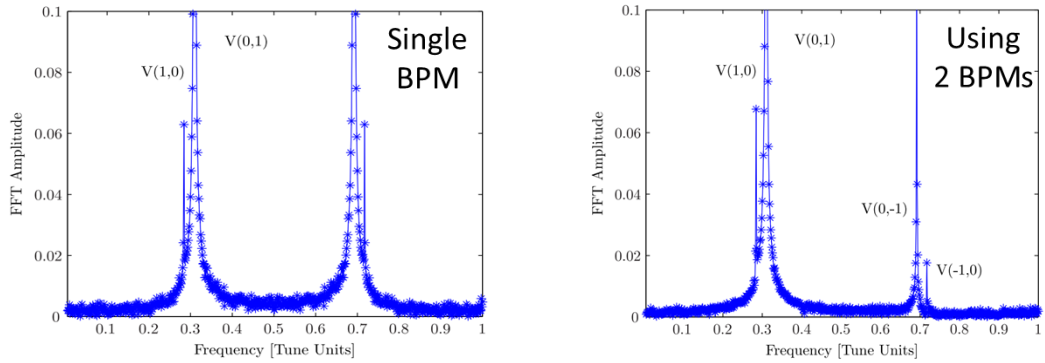

**Fig. 20:** Tune spectrum in a coupled machine using a single beam position monitor (left) and using a complex variable reconstructed from the data of two beam position monitors (right).

When using a single beam position monitor for such measurements, it is not possible to distinguish the phase of the oscillations. However, if two beam position monitors are used, then a complex variable can be created that reconstructs both amplitude and phase (Fig. 20). This technique can be applied to the whole machine, using turn-by-turn data from neighbouring beam position monitors, to determine the local coupling throughout the ring []. Such an example for the LHC is presented in Fig. 21, which also shows the result of applying coupling corrections on the local coupling throughout the machine. Such measurements are only possible with dedicated beams as they use the standard beam position system and therefore require strong excitation to obtain acceptable signals. A single pick-up, optimised for tune measurement, is much more sensitive and can therefore be used for continuous coupling measurement but can greatly over or under estimate the global coupling due to local variations.

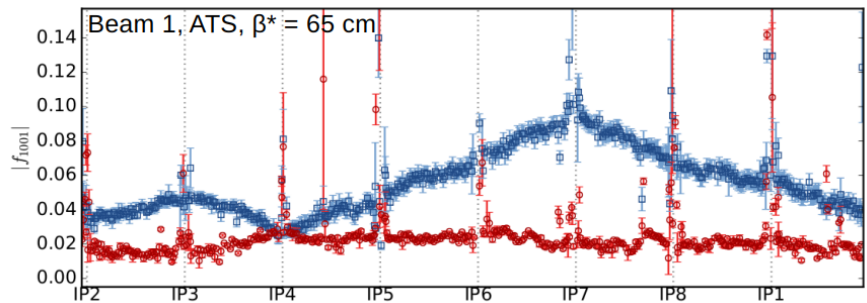

**Fig. 21:** Local coupling measurements throughout the LHC. Initial measurements in blue, with measurements after coupling optimisation in red.

## Acknowledgements

The author would like to thank M. Gasior for many of the figures reproduced in this contribution and M. Gasior and H. Schmickler for their input on the content.

## References

[1] M. Gasior and R. Jones, " The principle and first results of betatron tune measurement by direct diode detection", CERN-LHC-Project-Report-853 (2005).

[2] D. McGinnis, "Chromaticity Measurements Using Phase Modulated RF and Vector Signal Analyzers", FERMILAB-PBAR-NOTE-656 (2001).

[3] C.Y. Tan, "Chromaticity tracking with a phase modulation/demodulation technique in the Tevatron", *Nucl. Instrum. Methods Phys. Res. A.*, Vol 602, (2009), pp. 352-356.

[4] G. Jackson, "Tune Spectra in the Tevatron Collider", in Proc. 13th Particle Accelerator Conf. (PAC'89), Chicago, IL, USA, Mar. 1989, pp. 860-863.

[5] G. Rehm *et al.*, "Measurement of lattice parameters without visible disturbance to user beam at Diamond light source", in Proc. 14th Beam Instrumentation Workshop (BIW'10), Santa Fe, NM, USA, May 2010, paper MOCNB01, pp. 44-48.

[6] R. Singh *et al.*, Interpretation of transverse tune spectrum in a heavy-ion synchrotron", Phys. Rev. Acc. Beams 13, 034201 (2013).

[7] D. Boussard, "Schottky noise and beam transfer function diagnostics", CAS - CERN Accelerator School : 5th Advanced Accelerator Physics Course, Rhodes, Greece, 20 Sep - 1 Oct 1993, pp.749-782 (CERN-1995-006).

[8] A. Jansson, P. Lebrun, and R. Pasquinelli, "Experience with the 1.7 GHz Schottky Pick-ups in the Tevatron", in Proc. 9th European Particle Accelerator Conf. (EPAC'04), Lucerne, Switzerland, Jul. 2004, paper THPLT135.

[9] K. A. Brown, M. Blaskiewicz, C. Degen, and A. Della Penna, "Measuring transverse beam emittance using a 2.07 GHz movable Schottky cavity at the Relativistic Heavy Ion Collider", PhysRevSTAB, **12**, 012801 (2009).

[10] M. Betz, O.R. Jones, T. Lefevre, and M. Wendt, "Bunched-beam Schottky monitoring in the LHC", *Nucl. Instrum. Methods Phys. Res. A.*, Vol 874, (2017), pp. 113-126.

[11] D. Cocq, O.R. Jones and H. Schmickler, "The measurement of chromaticity via a head-tail phase shift", in Proc. 8th Beam Instrumentation Workshop (BIW'98), SLAC, USA, May 1998. CERN SL-98-062 BI.

[12] N. Catalan-Lasheras, S. Fartoukh, and R. Jones, "Recent advances in the measurement of chromaticity via head-tail phase shift analysis", in Proc. 6th European Workshop on Beam Diagnostics and Instrumentation for Particle Accelerators (DIPAC'03), Mainz, Germany, May 2003, paper PM08.

[13] A. Boudsko *et al.*, "Chromaticity measurements at Hera-p using the head-tail technique with chirp excitation", proceedings of the 4th European Workshop on Diagnostics and Instrumentation for Particle Accelerators (DIPAC'99), Chester, UK, May 1999.

[14] C.-Y. Tan and V. H. Ranjbar, "Chromaticity measurement using a continuous head-tail kicking technique", in Proc. 22nd Particle Accelerator Conf. (PAC'07), Albuquerque, NM, USA, Jun. 2007, paper FRPMS014, pp. 3916-3918.

[15] S. Fartoukh, "A theory of the Beam Transfer Function (BTF) with amplitude detuning and chromaticity induced head-tail phase shift", LHC Project Report 986.

[16] R. Jones, P. Cameron, and Y. Luo, "Towards a Robust Phase Locked Loop Tune Feedback System", BNL-C-A/AP/204 (2005).

[17] T. Persson and R. Tomás, "Improved control of the betatron coupling in the Large Hadron Collider", Phys. Rev. Spec. Top. Accel. Beams 17 (2014) 051004.